\documentclass[a4paper]{article}
\pdfoutput=1
\usepackage{jheppub}

\renewcommand{\O}[1]{\ensuremath{\mathcal{O}\left(#1\right)}}
\renewcommand{\H}{\ensuremath{\mathcal{H}}}
\newcommand{\bt}{\ensuremath{\tilde{b}}}
\newcommand{\bh}{\ensuremath{\hat{b}}}
\newcommand{\ket}[1]{\ensuremath{\left|#1\right\rangle}}
\newcommand{\bra}[1]{\ensuremath{\left\langle#1\right|}}
\newcommand{\vev}[1]{\ensuremath{\left\langle#1\right\rangle}}
\newcommand{\tr}{\ensuremath{\mathrm{Tr}}}

\newcommand{\firewalls}{Bousso:2012as,Nomura:2012sw,Mathur:2012jk,Chowdhury:2012vd,Susskind:2012rm,Banks:2012nn,Ori:2012jx,Bena:2012zi,Giveon:2012kp,Brustein:2012jn,Susskind:2012uw,Hossenfelder:2012mr,Avery:2012tf,Nomura:2012cx,Hwang:2012nn,Rama:2012fm,Nomura:2012ex,Giddings:2012gc,Larjo:2012jt,Papadodimas:2012aq,Saravani:2012is,Jacobson:2012gh,Harlow:2013tf,Susskind:2013tg,Kim:2013fv,Avery:2013exa,Chowdhury:2013tza,Lowe:2013zxa,Verlinde:2013uja,Maldacena:2013xja,Mathur:2013gua,Page:2013mqa,VR}

\title{Still No Rindler Firewalls}

\author{Michael Gary}

\affiliation{
  Institute for Theoretical Physics\\
  Vienna University of Technology\\
  Wiedner Hauptstra{\ss}e 8--10/136\\
  1040 Vienna, Austria
}

\emailAdd{
  mgary@hep.itp.tuwien.ac.at
}

\abstract{
  There has been much discussion on the possibility of firewalls at the horizon-scale in black hole physics, including questions regarding the presence or absence of firewalls at apparent horizons, such as the Rindler horizon and the horizon of the Poincar\'{e} patch of Anti-de Sitter space. We argue against the presence of such apparent firewalls by demonstrating that one recent argument for firewalls in black holes does not extend to these cases. We also include some brief remarks on some claims in the recent firewall literature. 
}


\begin{document}
\maketitle
\flushbottom

\section{Introduction}

It has long been hoped that black holes might provide a unique proving ground for ideas of Quantum Gravity, since within a black hole horizon, classical gravity breaks down in a singularity. The semi-classical calculation demonstrating that black holes evaporate provides a concrete realization of this hope, in that Unitarity and locality are brought into sharp conflict with each other. It is clear that there must be a false assumption somewhere in the calculation, as it leads to a paradox, but what is particularly troubling about the calculation is that effective field theory has broken down without any internal indication of its failure. After all, much of what makes effective field theory so useful is that it typically provides a warning when it is no longer applicable. Nonetheless, in the case of black hole physics, no warning sign is visible and we are left with contradictory results---all calculations take place in regions of low curvature, and since nothing is special about the horizon from the perspective of a local observer, effective field theory should be perfectly valid. 

With the development of a number of new tools, such as the AdS/CFT correspondence and holographic entanglement entropy, we ought to have more of a handle on the situation. To some extent, we do, in that we have good reason to believe that it is Unitarity which survives in Quantum Gravity, but in other aspects, we remain as confused as ever. In particular, we are still left with a breakdown of effective field theory without any internal indication. In fact, these new tools have simply sharpened the problem, as they have led to the current firestorm surrounding the controversial claims of \cite{AMPS,AMPSS} that the in-falling observer experiences interesting Planck-scale physics at the horizon \cite{\firewalls}. 

In section \ref{firewalls}, we explain the argument for the presence of firewalls at the black hole horizon, focusing on the new argument put forward by Marolf and Polchinski \cite{Marolf:2013dba}. We then move on in section \ref{geometry} to explain the geometry of Rindler and Poincar\'{e} patches of AdS and demonstrate how to evade this new argument for firewalls in the case of these horizons. Finally, we conclude and make some brief remarks about some claims that have been made in the recent firewall literature in \ref{concl}.

\section{Firewalls for AdS Black Holes}
\label{firewalls}

In some sense, the sharpest version of the firewall problem occurs in AdS, where we ought to have complete control over the physics through the CFT. Here I will review the most recent argument for the existence of firewalls at the horizon of AdS black holes formed by collapse as outlined in \cite{Marolf:2013dba}.

The Hilbert space of gravity in AdS$_D$, or equivalently of the CFT living on the boundary of AdS$_D$, $\H$, is spanned by states created by the product of insertions of local operators on the boundary of AdS in the past. It is important that we are allowed to insert operators anywhere on a sufficiently large region of the boundary or we will not be able to form a complete basis for $\H$. One example of a sufficiently large boundary region was given in \cite{Gary:2011kk}, where an explicit construction of a complete set of massless states was given using operator insertions a boundary region given by $S^{D-2}\times\left(-\pi,0\right)$. For our purposes, it will be sufficient to use this set of operators as a (possibly over-complete) basis for $\H$. Furthermore, if we insert operators with sufficiently high energy, these states will collapse to form large black holes in the bulk. 

Moving to a bulk description in effective field theory, let us label modes which are outside the horizon by $b$, modes inside the horizon by $\bt$, and modes adapted to an in-falling observer, which are smooth across the horizon, by $a$. The modes $b$ have an image in the boundary CFT given by $\bh$, which can be determined perturbatively in a $1/N$ expansion, with the leading component given in \cite{Banks:1998dd,Balasubramanian:1998de} and the subleading corrections determined using the equations of motion in \cite{Kabat:2011rz}. The operators $\bh$ are supported in a region on the boundary of AdS after the sources are sent in to form a black hole but before the horizon reaches the boundary. In particular, if we imagine sending in a shell of matter to form a black hole at time $t=0$, the horizon does not reach the boundary until an AdS time $\pi$ after the source is sent inward from the boundary. Thus, the region of support for $\bh$ is $S^{D-2}\times(0,\pi)$, which as we already have argued is sufficient to form a basis for $\H$.

Let us momentarily consider the argument that leads to Hawking radiation. If we assume the horizon is a smooth place, then the in-falling observer should be in the vacuum state, at least with regards to modes $a$ that are centered around high frequencies $\omega$, and thus
\begin{equation}
\vev{N_{a(\omega)}}=\vev{a^\dagger a}=0\ .
\end{equation}
Performing a Bogoliubov transformation to find what this corresponds to in terms of the $b$ modes, 
\begin{equation}
b=Ba + Ca^\dagger\ ,
\end{equation}
we find a thermal spectrum for the outside observer $b$. 

Returning to our description in terms of the dual CFT, since the states $\bh$ span $\H$, we can consider a basis of states $\left\{\ket{\psi}\right\}$ in which $N_{\bh}=\bh^\dagger\bh$ is diagonal. In this basis, because $N_{\bh}$ is thermal in the in-falling vacuum, 
\begin{equation}
\bra{\psi}\hat{N}_a\ket{\psi}\geq\O{1}\ , 
\end{equation}
where we have assumed that there is an image of $N_a$ in the boundary CFT, $\hat{N}_a$. Averaging over the Hilbert space, we find
\begin{equation}
\label{trace}
\frac{\tr_{\H}\left(\hat{N}_a\right)}{\tr_{\H}\left(1\right)}\geq\O{1}\ .
\end{equation}
This is startling, in that in a typical state of the Hilbert space, rather than seeing the vacuum, the in-falling observer sees many excited modes. It is the presence of these excited modes for the typical in-falling observer that constitute the firewall.

\section{Rindler and Poincar\'{e} Horizons}
\label{geometry}

Now that we have understood the presence of a firewall at black hole horizons, let us attempt to understand whether other types of horizons have similar firewalls. In particular, we might now be concerned that this argument also applies to Rindler horizons, meaning that there should be firewalls in flat space, or at least in empty AdS. Such firewalls would be severely disconcerting at the least. 

\begin{figure}[tbp]
\centering
\includegraphics[width=0.45\textwidth]{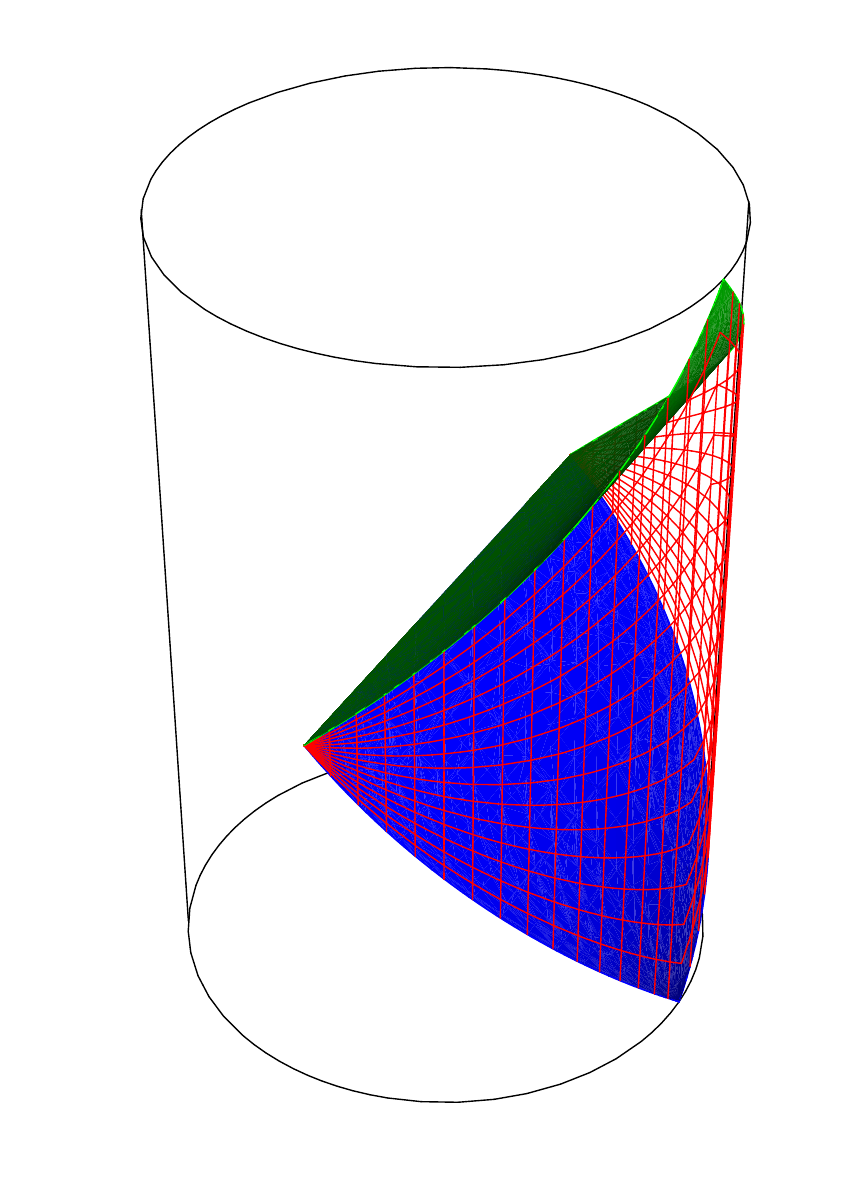}
\hfill
\includegraphics[width=0.45\textwidth]{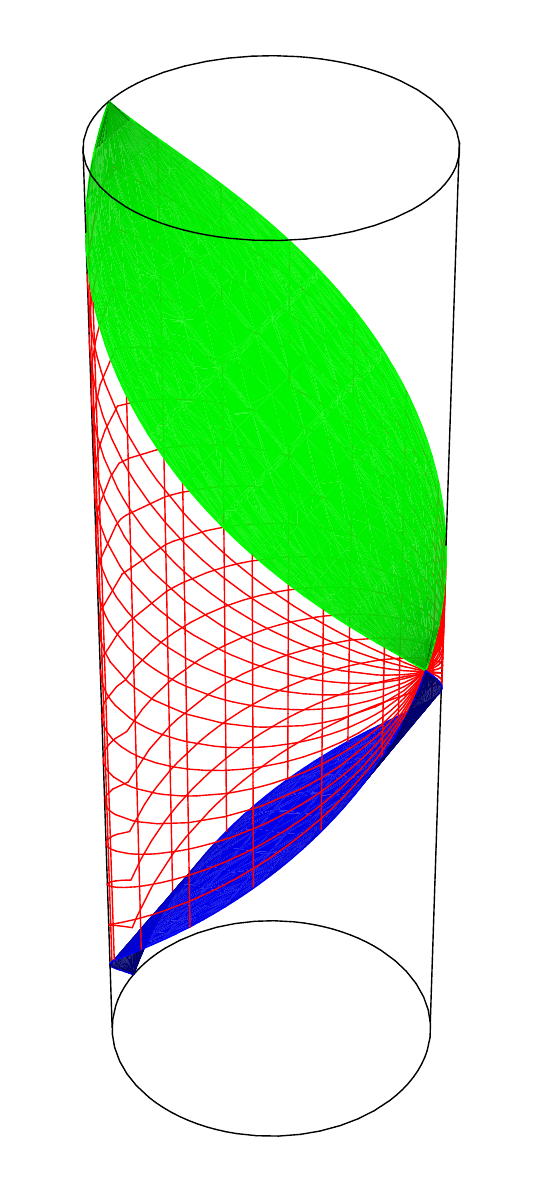}
\caption{\label{AdSPatches} The Rindler patch of AdS is shown at left and the Poincar\'{e} patch of AdS is shown at right. The past horizons of these patches are shown in blue, the future horizons in green, and the boundary in red.}
\end{figure}

We will first consider the Rindler patch of AdS, shown on the left in figure \ref{AdSPatches}. As in the Rindler wedge of Minkowski space, the Rindler patch covers the region of AdS accessible to an observer undergoing constant acceleration. We will label the modes in the accelerated observers frame by $b$, in analogy to the modes of the outside observer in the case of an AdS black hole. Similarly, the modes of an unaccelerated observer, who simply passes through the Rindler horizon, will be labeled by $a$. Once again, we can map the $b$ modes to their CFT image on the boundary of AdS, $\bh$, although this time the CFT operators are only supported in the wedge-shaped region of the boundary shown in red in figure \ref{AdSPatches}. 

While we can again consider a basis $\left\{\ket{\psi}\right\}$ for $\H$ in which $N_{\bh}$ is diagonal, there is a crucial difference from the black hole case. In particular, the states $\bh$ do not span $\H$. Because of this fact, the eigenspaces of $N_{\bh}$ are highly degenerate, which is vastly different from the case of a black hole. While it is possible to argue this from the bulk perspective, as has been done in \cite{Bousso:2012mh}, here we will give a purely boundary argument. If we take the Rindler wedge to be centered about the north pole of $S^{D-2}$ at time $t=0$. The boundary of the Rindler wedge $\partial R$ only covers half of the sphere at its widest point at $t=0$ and only ranges over $t\in(-\frac{\pi}{2},\frac{\pi}{2})$. If we insert an operator $\hat{\mathcal{O}}$ at the south pole at time $t=0$, by causality there is no operator with support in $\partial R$ that can detect the presence of $\hat{\mathcal{O}}$. Thus, it is clear that the operators $\bh$, with support only in $\partial R$, cannot generate a complete basis of $\H$. 

Because the operators $\bh$ do not generate a basis of $\H$, we cannot assert that the average over the Hilbert space in equation \ref{trace} is $\O{1}$ or larger. In other words, we are unable to argue that in a generic state in $\H$ the unaccelerated observer passing through the Rindler horizon would see many excited modes, and thus the argument of section \ref{firewalls} does not extend to producing firewalls for Rindler horizons. 

While this argument does not exclude the possibility of a firewall, there is no contradiction in the absence of a Rindler firewall, as in there is in the case of a black hole. Furthermore, it seems clear that if we do not turn on any operators outside of $\partial R$, the unaccelerated observer should pass through the horizon unmolested. More colloquially, we can choose to have a band of evil archers lying in wait behind the horizon to annihilate any observer who passes through, but unlike the case of the black hole horizon, we are under no obligation to do so. 

This argument also seems to extend to the Poincar\'{e} patch of AdS, shown to the right in figure \ref{AdSPatches}, although this case is more subtle, since at time $t=0$, only the south pole of $S^{D-2}$ is not included in the boundary region. Nonetheless, the states $\bh$ with support only in the boundary of the Poincar\'{e} patch do not span $\H$ by the same argument as before. The subtlety arises when considering the average over $\H$. If we call the subspace spanned by the states generated by $\bh$ $\H_p$ and the transverse space $\H_\perp$, if $\H_\perp$ is of measure zero in $\H$, then equation \ref{trace} will still hold, and we will once again be forced to accept a firewall.

\section{Conclusions}
\label{concl}

We have demonstrated that the recent argument put forward in favor of black hole firewalls in AdS does not extend to Rindler horizons. This addressed a major concern of the author and others, since the presence of a firewall in Rindler space would seem to contradict our everyday experience, which is thankfully free of firewalls. While our argument does not actually exclude the possibility of Rindler firewalls, as argued for in \cite{Czech:2012be}, among other places, it does demonstrate that they are not strictly necessary. In particular, while a typical state which is a product of arbitrary states on the left and right Rindler wedges will have a firewall at the Rindler horizon, it is not clear that this is the appropriate basis in which to consider the picture. In fact, such states do not have bounded total Minkowski/AdS energy and could not evolve from finite energy initial conditions, so there is a strong argument that this is not a good basis for considering the problem of an observer passing through a Rindler horizon.\footnote{I would like to thank D.~Harlow for clarifying this point.}

At this point, I would also like to include a few remarks on some recent works on the firewall paradox. Many of the arguments for firewalls rely on statements about generic states in the Hilbert space of quantum gravity. While the example of a one-sided black hole coupled to an external CFT evolving to a two-sided black hole \cite{VR} is intriguing, as is noted within the paper, the initial state and coupling seem to be so fine tuned as to disallow any statements about the genericity of the process. It is precisely the appearance of a firewall in the generic black hole microstate, not its absence from a specific finely tuned microstate, that leads to the paradox, and thus the argument of \cite{VR} is not sufficient to demonstrate the non-existence of firewalls at late times or to invalidate the analysis of \cite{AMPS,AMPSS}.

\acknowledgments

I would like to thank Daniel Harlow, Nima Lashkari, and Joe Polchinski for many puzzling discussions on the subject of firewalls and help in clarifying my thoughts. This work was supported in part by project I~1030-N27 of the Austrian Science Fund (FWF). Additionally, I would like to thank the Kavli IPMU for their hospitality while this work was completed.

\end{document}